\def\year{2021}\relax
\documentclass[letterpaper]{article} 
\usepackage{aaai21}  
\usepackage{times}  
\usepackage{helvet} 
\usepackage{courier}  
\usepackage[hyphens]{url}  
\usepackage{graphicx} 
\usepackage{natbib}  
\usepackage{caption} 
\usepackage{amsmath}
\usepackage{amssymb}
\usepackage{bm}
\usepackage{amsthm}
\usepackage{algorithmicx}
\usepackage{subcaption}
\usepackage{booktabs}
\DeclareMathOperator*{\E}{\mathbb{E}}

\title{Clinical trial of an AI-augmented intervention for HIV prevention in youth experiencing homelessness}

\author {
Bryan Wilder\textsuperscript{\rm 1}, Laura Onasch-Vera\textsuperscript{\rm 2}, Graham Diguiseppi\textsuperscript{\rm 2}, Robin Petering\textsuperscript{\rm 3},\\Chyna Hill\textsuperscript{\rm 2}, Amulya Yadav\textsuperscript{\rm 4}, Eric Rice\textsuperscript{\rm 2}, Milind Tambe\textsuperscript{\rm 1}\\
}
\affiliations {
    \textsuperscript{\rm 1} Center for Research on Computation and Society, \\John A. Paulson School of Engineering and Applied Sciences, Harvard University \\
    \textsuperscript{\rm 2} Suzanne Dworak-Peck School of Social Work and Center for Artificial Intelligence in Society,\\ University of Southern California \\
    \textsuperscript{\rm 3} Lens Co \\
    \textsuperscript{\rm 4} College of Information Sciences and Technology, Pennsylvania State University \\
}
\begin{document}

\maketitle

\begin{abstract}
Youth experiencing homelessness (YEH) are subject to substantially greater risk of HIV infection, compounded both by their lack of access to stable housing and the disproportionate representation of youth of marginalized racial, ethnic, and gender identity groups among YEH. A key goal for health equity is to improve adoption of protective behaviors in this population. One promising strategy for intervention is to recruit peer leaders from the population of YEH to promote behaviors such as condom usage and regular HIV testing to their social contacts. This raises a computational question: which youth should be selected as peer leaders to maximize the overall impact of the intervention? We developed an artificial intelligence system to optimize such social network interventions in a community health setting. We conducted a clinical trial enrolling 713 YEH at drop-in centers in a large US city. The clinical trial compared interventions planned with the algorithm to those where the highest-degree nodes in the youths' social network were recruited as peer leaders (the standard method in public health) and to an observation-only control group. Results from the clinical trial show that youth in the AI group experience statistically significant reductions in key risk behaviors for HIV transmission, while those in the other groups do not. This provides, to our knowledge, the first empirical validation of the usage of AI methods to optimize social network interventions for health. We conclude by discussing lessons learned over the course of the project which may inform future attempts to use AI in community-level interventions.
\end{abstract}

\section{Introduction}

Each year, approximately 4.2 million youth in the United States experience some form of homelessness \cite{morton2018prevalence}. One of the key health challenges for this population is high HIV prevalence, with reported prevalence in the range of 2-11\% \cite{young2011online}, up to 10 times the rate for youth with access to stable housing \cite{nhch2012}.


One proposed mechanism for fostering behavior change in high-risk populations is the \textit{peer change agent} model. The main idea is to recruit peer leaders from the population of youth experiencing homelessness (YEH) to serve as advocates for HIV awareness and prevention. Use of peer leaders has been suggested in the public health and social science literature due to the central role that peers play in risk behaviors for YEH, including related to HIV spread \cite{green2013shared,rice2010internet,doi:10.2105/AJPH.2011.300295}. Indeed, peer change agent models have succeeded in past HIV prevention interventions in other contexts \cite{medley2009effectiveness}. However, there have also been notable failures \cite{nimh2010results}, and it has been argued such failures may be attributable to how peer leaders are selected \cite{schneider2015new}. The long-standing and most widely adopted method in the public health literature for selecting peer leaders is to identify the most popular individuals in the social network of the youth \cite{kelly1997randomised} (formally, the highest degree nodes). This poses the question: are high-degree youth the best peer leaders to disseminate messages about HIV prevention? This question has relevance far beyond HIV prevention; analogous social network interventions are used widely across development, medicine, education, etc.\ \cite{kim2015social,paluck2016changing,banerjee2013diffusion,valente2007identifying}. 

Information dissemination on social networks is the focus of a long line of research in computer science. In particular, the \textit{influence maximization} problem, formalized by \cite{kempe2003maximizing}, asks how a limited number of seed nodes can be selected from a social network to maximize information diffusion. Influence maximization has been the subject of extensive work by the theoretical computer science and artificial intelligence communities \cite{chen2009efficient,chen2010scalable,goyal2011celf++,borgs2014maximizing,tang2014influence}. However, to our knowledge, no work prior to this project had connected the computational literature on influence maximization to the use of network-driven interventions in public health and related fields. Computational work has mainly focused on developing highly efficient algorithms for use on large-scale social media networks (often motivated by advertising), while interventionists in health domains have not used explicitly algorithmic approaches to optimize the selection of peer leaders. Previous computational work assumed access to data (e.g., the full network structure and a model of information spread) which are simply not available in a public health context.

This paper reports the results of a project which bridges the gap between computation and health interventions. As a research team composed of computer scientists and social workers, we developed, implemented, and evaluated an intervention for HIV prevention in YEH where the peer leaders are algorithmically selected. This intervention was developed over the course of several years, alternating between algorithm design and smaller-scale pilot tests to evaluate feasibility. The final system, which we refer to as CHANGE (CompreHensive Adaptive Network samplinG for social influencE), was evaluated in a large-scale clinical trial enrolling 713 youth across two years and three sites. The trial compared interventions planned with CHANGE to those using the standard public health methodology of selecting the youth with highest degree centrality (DC), as well as an observation-only control group (OBS). \textit{Results from this clinical trial demonstrate that CHANGE was substantially more effective than the standard DC method at increasing adoption of behaviors protective against HIV spread.} To our knowledge, this is the first empirically validated success of using AI methods to improve social network interventions for health. It is critically important for ``AI for Social Good" work to result in deployed and rigorously evaluated interventions, and this paper provides one such example.

The remainder of the paper is organized as follows. First, we survey related work from both a computational and application perspective. Second, we introduce a formalization of the problem of selecting peer leaders from a computational perspective. Third, we briefly review the design of the CHANGE system to address this problem (deferring most details to earlier technical publications \cite{wilder2018end,wilder2018equilibrium,wilder2018maximizing}). Fourth, we present the design of the clinical trial. Fifth, we present and analyze results from the trial. Sixth, we discuss lessons learned over the course of the project which may help inform future attempts to design and implement AI-augmented public health interventions. 

\section{Related work}

A great deal of research in computer science has been devoted to the influence maximization problem. The majority of this has focused on computationally efficient algorithms for large networks \cite{chen2009efficient,chen2010scalable,goyal2011celf++,borgs2014maximizing,tang2014influence} and assumes that the underlying social network and model of information diffusion are perfectly known. There is also more recent literature on algorithms to learn or explore these properties. Predominantly though, such work requires many repeated interactions with the system. For example, algorithms to estimate the parameters of an unknown model of information diffusion \cite{du2014influence,pouget2015inferring,narasimhan2015learnability,he2016learning,kalimeris2018learning} typically require the observation of hundreds of cascades on the same network. Collecting this amount of data is intractable for public health interventions, where a single round of the intervention takes months. Other work concerns the bandit setting, where the algorithm can repeatedly select sets of nodes and observe the resulting cascade \cite{wen2017online,chen2013combinatorial,wang2017improving}. Similarly, these algorithms accept poor performance in early rounds as the price for improvement over the long run, but waiting tens or hundreds of rounds for improved performance is not an option in our domain. Such techniques are a much better fit for problems concerning online social networks (for example, in advertising domains) where repeated experiments and large datasets are possible.

The most closely related related computational work to ours concerns a robust version of the influence maximization problem \cite{he2018stability,chen2016robust,lowalekar2016robust}, building on the earlier work of \cite{krause2008robust} on general robust submodular maximization problems. Our algorithm for robust submodular optimization, for which an overview is provided below, differs from these approaches mainly in that it solves a fractional relaxation of the problem instead of repeatedly calling a greedy algorithm for discrete submodular optimization, which helps improve computational performance. 

There is a large literature on social network interventions in public health \cite{valente2007identifying,kim2015social}, clinical medicine \cite{young2003role}, international development \cite{cai2015social,banerjee2013diffusion}, education \cite{paluck2016changing}, etc. Common strategies involve selecting high degree nodes (as compared to in our trial), selecting nodes at random, or asking members of the population to nominate others as influencers. The empirical evidence for the relative effectiveness of different strategies is mixed; \cite{kim2015social} reports no or marginal improvement for nominations vs random selections (depending on the outcome measure), while \cite{banerjee2019using} report statistically significant improvements for a nomination-based selection mechanism. \cite{chin2018evaluating} introduce improved statistical methods to compare the effectiveness of seeding strategies and conclude that nomination-based strategies do not measurably improve performance. Indeed, \cite{akbarpour2018diffusion} show that in some theoretical network models it may be preferable to recruit a slightly larger number of influencers at random rather than carefully map the network. We contribute to this literature by developing and empirically evaluating an algorithmic framework which combines both features reminiscent of the nomination-based strategies proposed by others (for gathering information about network structure) as well as robust optimization techniques for jointly optimizing the entire set of influencers who are selected (not part of previous empirically evaluated strategies). Our clinical trial demonstrates statistically significant improvements from this strategy compared to the baseline of selecting high-degree nodes, providing (to our knowledge) the first real-world evidence that systematic optimization leads to improved results.

\section{Problem description}

The population of youth are the nodes of a graph $G = (V, E)$. We seek to recruit a set of youth $S$ to be peer leaders, where $|S| \leq k$. In domain terms, this budget constraint reflects the fact that peer leaders are given a resource-intensive training and support process. The objective is to maximize the total expected number of youth who receive information about HIV prevention, given by the function $f(S)$. Here, $f$ encapsulates the dynamics of a probabilistic model of information diffusion across the network (discussed below). The optimization problem $\max_{|S| \leq k} f(S)$ is the subject of the well-known influence maximization problem. When the objective function $f$ is instantiated using common models for information diffusion, the resulting optimization problem is submodular (i.e., there are diminishing returns to selecting additional peer leaders). While finding an optimal solution is NP-hard, a simple greedy algorithm obtains a $\left(1 - 1/e\right)$-approximation \cite{kempe2003maximizing}.

The most common choice for the model of information diffusion is the independent cascade model. In this model, each node who receives information transmits it to each of their neighbors with probability $p$. All such events are independent. The process proceeds in discrete time steps where each newly informed node attempts to inform each of their neighbors, and concludes when there are no new activations. $f(S)$ calculates the number of nodes who receive information when the nodes $S$ are informed at the start of the process, in expectation over the random propagation.

The standard influence maximization problem concludes here. However, while developing an algorithmic framework applicable to public health contexts, we came across challenges which must be solved before, during, and after the setting imagined in standard influence maximization. These challenges opened up new algorithmic questions, addressed in a series of publications in the AI literature \cite{wilder2018end,wilder2018equilibrium,wilder2018maximizing}. Here, we detail three steps for deploying an influence maximization intervention in the field.

First, information about the network structure $G$ must be gathered. Previous work on influence maximization assumed that the network structure is known in advance. While this assumption may be reasonable for online social networks, we aim to disseminate information through the network consisting of real-world interactions between youth at a given center. Moreover, pilot studies revealed that information from an online social network (Facebook) was a poor proxy for actual connections at the center -- not all youth used Facebook, and of those who did, many were not friends with their actual contacts at the drop-in center. Instead, network information must be gathered through in-person interviews where social workers ask youth to list those who they regularly interact with. Collecting data in this manner is time-consuming and expensive, often requiring a week or more of effort on the part of the social work team. Accordingly, the first stage of our algorithmic problem is to decide which nodes to query for network information. The algorithm is allowed to make $M$ queries, where each query reveals the edges associated with the selected node. The queries can be adaptive, i.e., the choice of the $i$th node to be queried can depend on the answers given by nodes $1...i-1$.

Second, this network information is used to select an initial set of peer leaders. This stage more closely resembles the standard influence maximization problem. However, there is an additional complication that the propagation probability $p$ is not known. Indeed, there is no data source from which it could be inferred (as opposed to online platforms with abundant data; see related work). Instead, we formulate an uncertainty set $\mathcal{U}$ containing a set of possible values for $p$ which are consistent with prior knowledge (in CHANGE, we took $\mathcal{U}$ to be a discretization of the interval [0,1], reflecting limited prior knowledge). The aim is to find a set $S$ which performs near-optimally for every scenario contained in $\mathcal{U}$. Formally, this corresponds to the robust optimization problem 
\begin{align*}
    \max_{|S| \leq k} \min_{p \in \mathcal{U}} \frac{f(S, p)}{OPT(p)}
\end{align*}
where $OPT(p)$ denotes $\max_{|S| \leq k} f(S, p)$, i.e., the best achievable objective value if the propagation probability $p$ were known. Normalizing by $OPT(p)$ encourages the algorithm to find a set $S$ which simultaneously well-approximates the optimal value for each $p \in \mathcal{U}$ and avoids the trivial solution where solution to the inner $\min$ problem is always the smallest possible value of $p$. Note that since $OPT(p)$ is constant with respect to $S$, $\frac{f(S, p)}{OPT(p)}$ remains submodular with respect to $S$. Robust optimization of submodular functions is substantially more difficult than optimization of a single submodular function; in fact, it is provably inapproximable in general \cite{krause2008robust} and the aim is instead to approximate a tractable relaxation of the problem. 

Third, after an initial set of peer leaders $S$ is identified, recruitment proceeds in an adaptive manner. Not all youth invited to become peer leaders will actually attend the training session. A number of potential barriers exist, e.g., a given youth could have been arrested or not have had enough money for a bus ticket. Formally, we model that each youth who is invited will actually attend with probability $q$ (based on experience in pilot studies, we took $q = 0.5$), where the attendance of each youth is independent of the others. For a given value of $p$, the resulting objective function is $f(S, p, q)$, which takes an expectation over both the randomness in which nodes are successfully influenced at the start of the process and in the subsequent diffusion. It is easy to show \cite{wilder2018end} that $f$ remains submodular with this additional randomness. Because of this variation in attendance, as well as capacity limits for the initial training, peer leaders are recruited over multiple rounds, where the peer leaders selected in round $t$ can depend on those who were successfully recruited in rounds $1...t-1$. In each round $t$, we select a set of peer leaders $S_t$ with $|S_t| \leq k_t$ and observe which nodes are successfully recruited as peer leaders. The process continues for $T$ rounds in total.

\section{System design}

Our final proposed system for intervention planning is called CHANGE. CHANGE was originally introduced in \cite{wilder2018end}. The final version of CHANGE summarized here is nearly the same as the original, with the exception of the algorithm used for robust optimization, which was separately developed and published in \cite{wilder2018equilibrium}. We now provide an overview of CHANGE, mirroring the steps of the earlier problem formulation.

\subsubsection{Network sampling} CHANGE uses a simple but well-motivated heuristic to select a subset of nodes to be queried for network information (in the discussion section, we briefly review our earlier work on a more theoretically sophisticated solution, and the rationale for choosing a simpler method). The chosen method splits the query budget $M$ into two halves. Each query in the first half is made to a node selected uniformly at random from the network. Each query in the second half follows a query in the first half, and selects a uniformly random neighbor of the first node. This design is motivated by the friendship paradox, the observation that high-degree nodes are overrepresented when we sample random neighbors \cite{feld1991your}. Hence, the two stages of the query process balance between competing objectives: the first step encourages diversity, since random sampling ensures that we cover many different parts of the network, while the second step tends towards high-degree nodes who can reveal a great deal of network information. 

\subsubsection{Robust optimization} We now provide an overview of how CHANGE handles parameter uncertainty within a single stage of the planning process, before considering the multi-stage problem (with uncertain attendance) below. As mentioned above, max-min submodular optimization is NP-hard to approximate (within any nonzero factor) \cite{krause2008robust}. Accordingly, we need to somehow relax the problem to obtain meaningful guarantees. Let $\mathcal{I}$ denote the set of all feasible solutions (sets $S$ where $|S| \leq k$) and $\Delta(\mathcal{I})$ be the set of all distributions over $\mathcal{I}$ (i.e., the $|\mathcal{I}|$-dimensional simplex). We developed an algorithm for the problem
\begin{align} \label{problem:fractional}
    \max_{D \in \Delta(\mathcal{I})} \min_{p \in \mathcal{U}} \E_{S \sim D}\left[\frac{f(S, p)}{OPT(p)}\right]
\end{align}
which allows the algorithm to select a distribution over feasible sets and evaluates the worst case only in expectation over this distribution. In game theoretic terms, this allows the algorithm to select a mixed strategy instead of a pure strategy. At run-time, we sample from $D$; the resulting set has guaranteed performance in expectation over the sampling, but strong guarantees cannot be obtained ex-post for the sampled set (as a result of the computational hardness of the original max-min problem). However, in practice we find that sampling several random sets and selecting the best one gives excellent empirical performance (i.e., closely matching or exceeding the expected value of the distribution).

Our algorithm for this problem, detailed in \cite{wilder2018equilibrium}, uses a compact representation of the space of distributions (keeping track of only the marginal probability that each node is selected instead of each of the exponentially many potential subsets). It solves a fractional relaxation of the discrete max-min problem using this compact representation via a stochastic first-order method which is adapted to the particular properties of the objective. Then, we can use known rounding algorithms for submodular maximization to sample random sets from the distribution encoded by the solution to the fractional relaxation. This procedure guarantees a $(1 - 1/e)^2$-approximation for Problem \ref{problem:fractional}, which can be improved to $(1 - 1/e)$ with some additional steps (which we did not find empirically necessary). 

\subsubsection{Multi-stage intervention with attendance uncertainty} We handle the multi-stage nature of the intervention by running the robust optimization problem at each stage, calculating the objective function in expectation over which peer leaders will attend and conditioning on the selection of those who have attended previous interventions. Formally, this means that at stage $t > 1$, we solve \fontsize{9}{10}\selectfont
\begin{align*} 
    \max_{D \in \Delta(\mathcal{I})} \min_{p \in \mathcal{U}} \E_{S_t \sim D}\left[\frac{f(S_t \cup S_1 \cup ... \cup S_{t-1}, p, q)}{\max_{|S^*| \leq k} f(S^* \cup S_1 \cup ... \cup S_{t-1}, p, q)}\right]
\end{align*}\normalsize
where $S_1...S_{t-1}$ denote the sets of peer leaders who were succsesfully recruited in each previous stage. It is easy to show that the inner objective $f$ remains submodular in $S_t$ (see \cite{wilder2018end}), and so we retain the earlier guarantees on the quality of the solution obtained at each individual step. Moreoever, in \cite{wilder2018end} we show that the multi-stage problem as a whole enjoys the property of \textit{adaptive submodularity}, meaning that for any fixed parameter value $p$, solving  
\begin{align*}
    \max_{D \in \Delta(\mathcal{I})} \E_{S_t \sim D}\left[\frac{f(S_t \cup S_1 \cup ... \cup S_{t-1}, p, q)}{\max_{|S^*| \leq k} f(S^* \cup S_1 \cup ... \cup S_{t-1}, p, q)}\right]
\end{align*}
at each step $t$ and selecting the resulting set $S_t$ enjoys an approximation guarantee relative to the optimal adaptive policy for selecting a sequence of sets $S_1...S_t$ (again, with respect to a fixed $p$). More detailed discussion of the theoretical properties can be found in \cite{wilder2018end}. 

\section{Study design}

\begin{figure}
    \centering
    \includegraphics[width=3.2in]{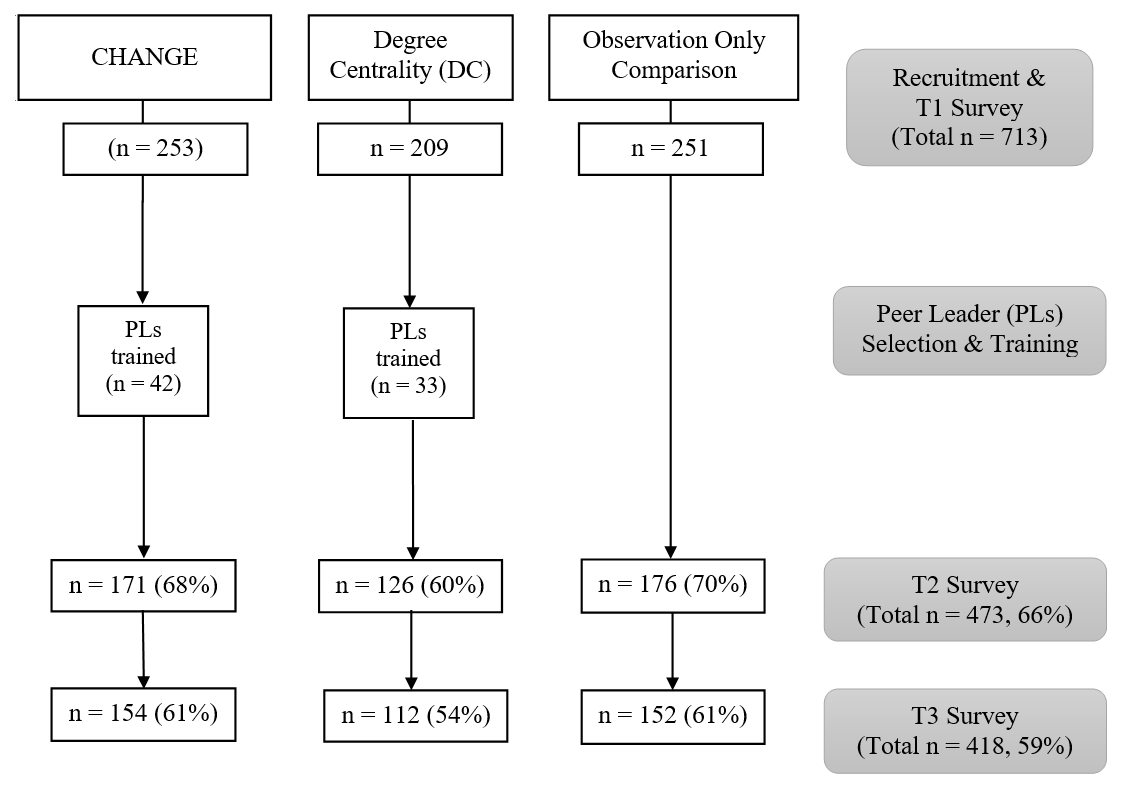}
    \caption{Number of participants recruited and retained in each arm of the study.}
    \label{fig:participants}
\end{figure}

We now move to the empirical portion of the project and provide an overview of the design of the clinical trial. All study procedures were approved by our institution's Institutional Review Board. The study was designed to compare the efficacy of two different means of selecting peer leaders: the CHANGE system described above and the standard DC approach in public health (selecting the highest-degree youth). We additionally included an observation-only control group (OBS), for three arms in total. The study was conducted at three drop-in centers for YEH in a large US city. Drop-in centers provide basic services to YEH (e.g., food, clothing, case management, mobile HIV testing). Due to high transience in the YEH population, most clients at a given center leave within approximately six months. Accordingly, we tested each of the three methods at each of the the three drop-in centers (giving nine deployments in total, each with a unique set of youth)\footnote{Randomizing treatments at an individual level is clearly impossible for an social network intervention, so this is an example of a quasi-experimental design where entire populations of youth were assigned to one treatment or another.}, ensuring that successive deployments at a given drop-in center were separated by six months. Youth were only allowed to enroll in the study once, so even the small number of youth who were present at the center across multiple deployments were included only on the first time they attempted to enroll. Testing each method at each drop-in center helps account for differences in the demographic and other characteristics of youth who tend to access services at each center.

Each of the nine deployments used the following procedure. Figure \ref{fig:participants} shows the number of youth recruited and retained for each phase of the study in each arm. 

First, youth were recruited at the drop-in center over the course of a week to participate in the study. All participants gave informed consent. Each participant completed a baseline survey which assessed demographic characteristics, sexual behaviors, and HIV knowledge. Demographic characteristics included age, birth sex, gender identity, race/ethnicity, and sexual orientation. Youth were also surveyed about their living situation and relationship status. 

Second, peer leaders were selected and trained (for the CHANGE and DC arms of the study). Each individual training consisted of approximately 4 youth and there were 3-4 trainings per deployment (depending on exact attendance). In total, approximately 15\% of survey participants in each deployment were trained as peer leaders. In the CHANGE arm of the study, network information was queried from approximately 20\% of the participants (sampled according to the mechanism described above). In the DC arm, we used a full survey of the network to find high-degree nodes, in order to give the strongest possible implementation to compare to.

Third, peer leaders had three months to disseminate HIV prevention messages. Peer leaders were supported via 7 weeks of 30-minute check-in sessions with study researchers, which focused on positive reinforcement of their successes as well as problem-solving strategies and goals for the future. All peer leaders attended at least one check-in session, with modal attendance at five sessions. Peer leaders received \$60 in compensation for attending the initial training and \$20 for each check-in session. 

Fourth, follow-up surveys were administered to the original study participants from the first step. Follow-up surveys assessed the same characteristics as the baseline survey. Differences in reported sexual behavior between baseline and follow-up were used as the primary metrics to evaluate the interventions. All such metrics were self-reported; we followed best practices in social science research to minimize bias in self-reported data (surveys were self-administered on a tablet and participants were guaranteed anonymity, each of which aim to reduce social desirability bias in reporting sensitive information). Additionally, any bias would be expected to influence each arm of the study equally, including the observation-only control group. 
 
 The training component of the peer change agent intervention was delivered by two or three facilitators from the social work research team. The training lasted approximately 4 hours (one half-day). Training was interactive and broken into six 45-minute modules on the mission of peer leaders (sexual health, HIV prevention, communication skills, leadership skills, and self-care). Peer leaders were asked to promote regular HIV testing and condom use through communication with their social ties at the drop-in center.

\section{Study results}

We now present the results of the clinical trial, starting with an overview of the outcome variables and methodology for statistical analysis, and then giving the main results. 

\subsection{Outcome variables}
We compare two outcome variables across arms of the study. First, condomless anal sex (CAS), assessed via a survey question asking whether youth had anal sex without a condom at least once in the previous month. Second, condomless vaginal sex (CVS), assessed via a survey question asking whether youth had vaginal sex without a condom at least once in the previous month. CAS and CVS are both important behavioral risk factors for HIV transmission and so provide a direct assessment of the success of the intervention at producing a material health impact.

\subsection{Statistical methodology}
We provide both the average value of each outcome variable at each time point for the three arms of the study as well as an analysis of statistical significance. The statistical analysis used a generalized estimating equations (GEE) model. GEE is an extension of generalized linear models which incorporates repeated measurements of data across a population. It is a standard choice for analysis of clinical data in this form \cite{gee}. We specified a linear model for each outcome variable which included terms for both the improvement caused by participating in a given arm of the study (our estimand of interest) as well as terms for a range of control variables which account for differences in demographics and the baseline rate of risk behaviors in each arm of the study. The demographic control variables were age, birth sex, transgender identity, LGBQ identity, the combination of male sex and LGBQ identity, race, committed relationship, housing status, and drop-in center. We also included a ``time" variable to account for changes in the entire population over time regardless of participation in a particular arm of the study. This combination of control variables helps separate the impact of the intervention from pre-existing differences between arms of the study and intervention-independent trends.

The linear model combined contributions from each of these variables through a logistic link function. Since each outcome is binary, we present results in the form of the odds ratio (OR), which measures the ratio in the odds of the outcome in youth who are exposed to a given intervention vs youth in the observation-only group (after controlling for demographics and baseline rate of risk behaviors). For all quantities, we also present 95\% confidence intervals and indicate where significant $p$-values are obtained. 

Results are known only for youth who completed the follow-up surveys, leading to missing data due to participant attrition (as is expected for a study enrolling YEH). Of the 713 participants who completed the baseline survey, 245 (34\%) missed the 1-month follow-up, 300 (42\%) missed the 3-month follow-up, and 180 (25\%) missed both follow-ups. However, missingness had no statistically significant association with CAS or CVS, indicating that youth were not significantly over or under represented in the follow-up data based on their baseline level of risk behavior.  
\newcommand{\ra}[1]{\renewcommand{\arraystretch}{#1}}
\begin{table}\centering
\ra{1.3}
\fontsize{8.5}{8.5}\selectfont
\begin{tabular}{@{}rccrrcccccccccc@{}}\toprule
& \multicolumn{2}{c}{CAS} & & \multicolumn{2}{c}{CVS} \\
\cmidrule{2-3} \cmidrule{5-6}   
& OR & CI && OR & CI\\ \midrule
\multicolumn{1}{l}{Baseline}\\
\multicolumn{1}{r}{CHANGE}& 1.43 & 0.91, 2.28 && 0.77 & 0.52, 1.13\\
\multicolumn{1}{r}{DC}& 1.49 & 0.89, 2.48 && 1.07 & 0.67, 1.68\\
\multicolumn{1}{l}{Post-intervention}\\
\multicolumn{1}{r}{CHANGE}& 0.69* & 0.49, 0.98 && 0.78$^\dagger$ & 0.57, 1.04\\
\multicolumn{1}{r}{DC}& 0.80 & 0.55, 1.17 && 0.88 & 0.62, 1.23 \\
\multicolumn{1}{l}{Time}& 1.05 & 0.82, 1.33 && 0.87 & 0.71, 1.06\\

\bottomrule
\end{tabular}
\caption{Results of statistical analysis. Each column gives the effect size and confidence interval for one of the outcome variables. Each row gives the corresponding estimates for one of the variables included in the GEE model. The ``baseline" category measures pre-existing differences between the groups (relative to the observation-only group) on enrollment in the study. The ``post-intervention" category measures the estimated impact of participating in each arm of the intervention (relative to the observation-only group, and after controlling for both demographics and baseline behaviors). ``Time" gives the estimated contribution of a trend over time independent of which arm of the study a participant was enrolled in. $^\dagger p < 0.1$; $^*p < 0.05$.} \label{table:results}
\end{table}
\subsection{Results}

We start by presenting the main results of the statistical analysis; the full results can be found in Table \ref{table:results}.

\subsubsection{CAS} We find that CAS reduced in the CHANGE group over time by a statistically significant amount (OR = 0.69, $p < 0.05$). The estimated OR of 0.69 indicates that, in the GEE estimates, a youth who is enrolled in the CHANGE arm of the study has 31\% lower odds to engage in CAS than if they were enrolled in the observation-only group. That is, a youth who is enrolled in CHANGE has 31\% lower odds to engage in CAS post-intervention than a youth with identical starting characteristics (including baseline rate of CAS) who did not receive the intervention. For the DC group, there was not a statistically significant change in CAS over time relative to the observation-only group. 

\subsubsection{CVS} The GEE model estimated that CVS decreased by a marginally statistically significant amount in the CHANGE group (OR = 0.78, $p < 0.1$). For the DC group, there was no statistically significant change in CVS over time relative to the observation-only group.

\begin{figure}
    \centering
    \includegraphics[width=1.5in]{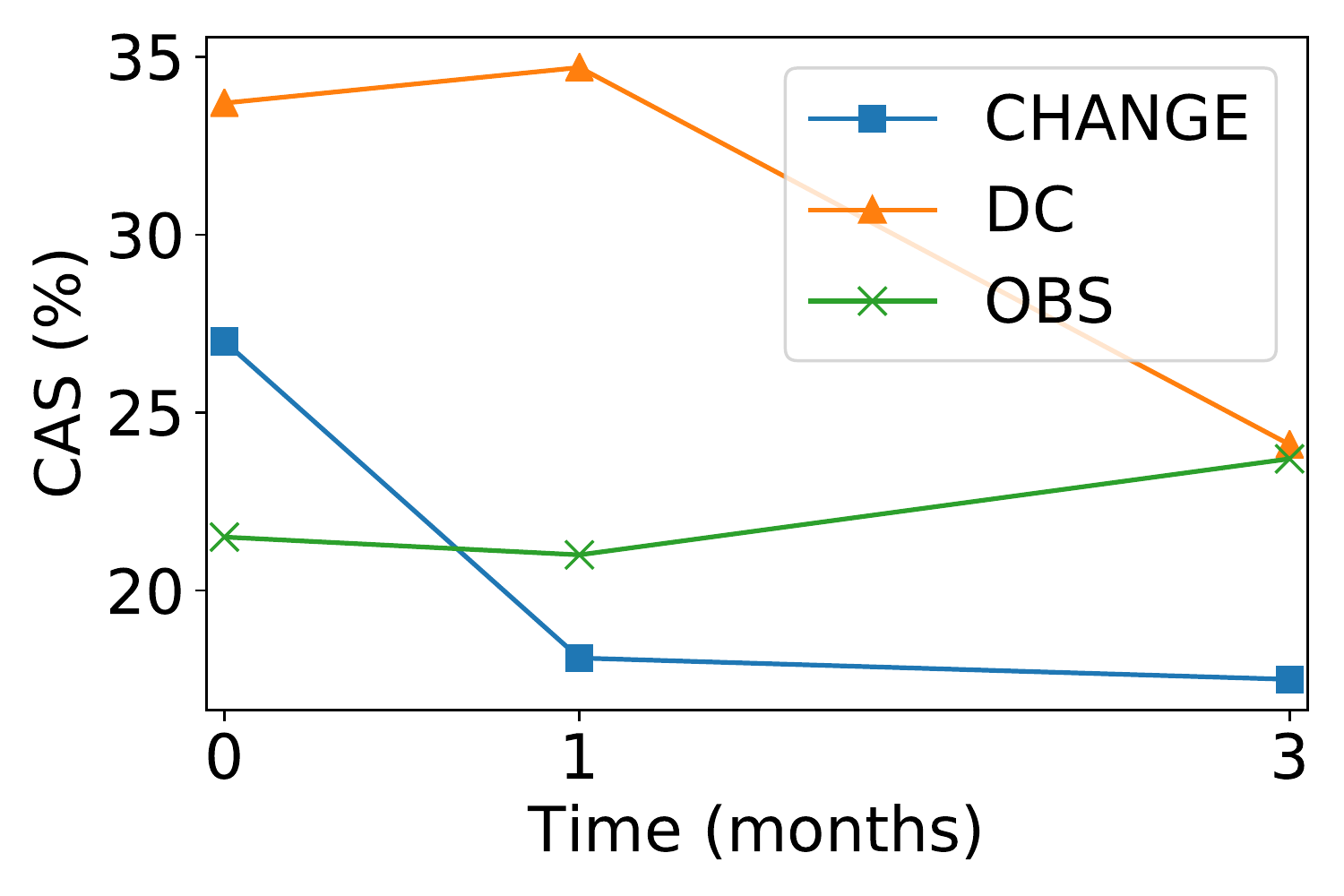}
    \includegraphics[width=1.5in]{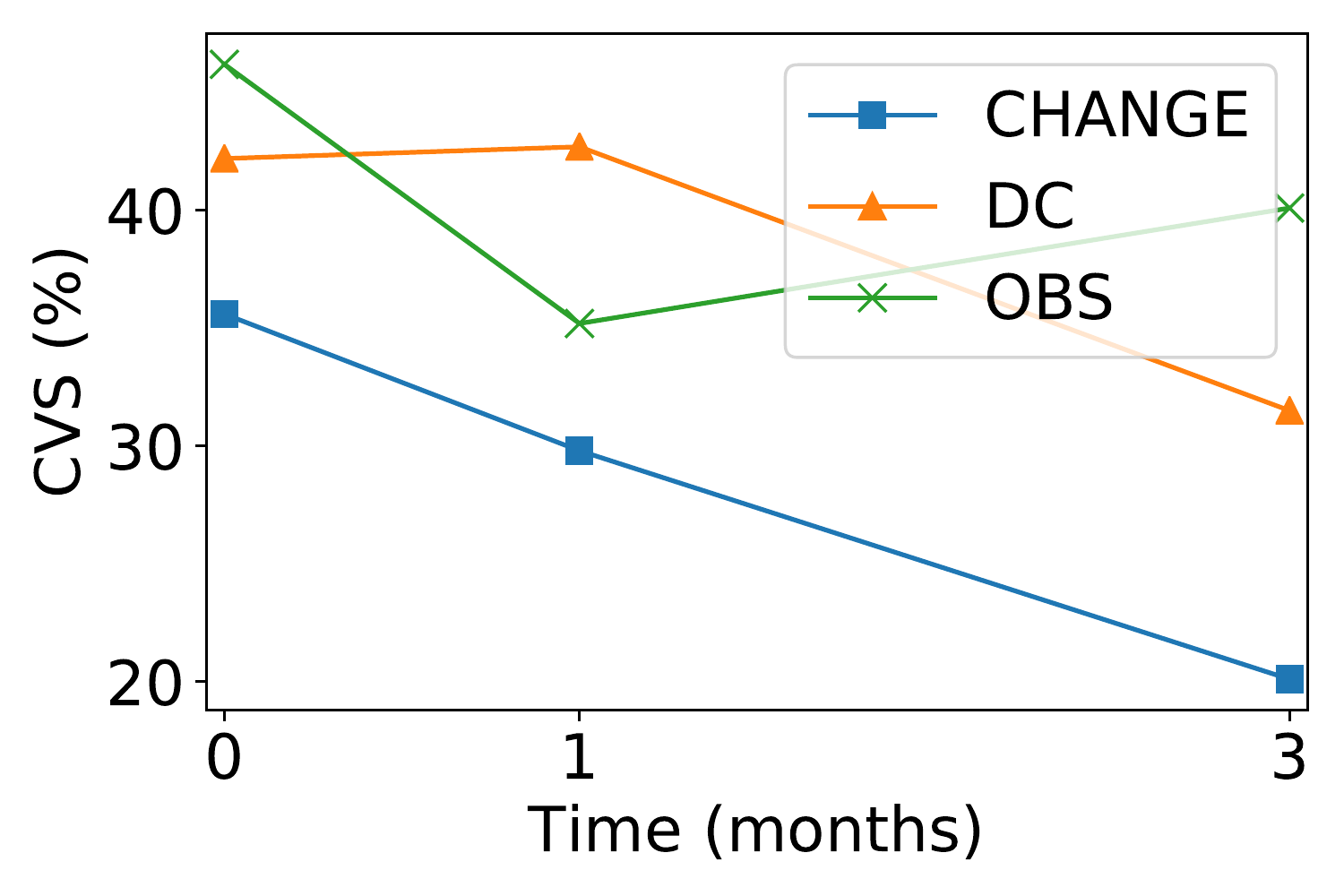}
    \caption{Average value of each outcome variable at each point in time for the three arms. These plots show the results without any statistical processing, while the analysis above attempts to control for pre-existing differences between participants in each arm.}
    \label{fig:results}
\end{figure}

We conclude from the analysis that only CHANGE provided a statistically significant improvement in HIV risk behaviors compared to the observation-only baseline. 

Direct examination of the average values of the outcome variables for each arm at each point in time (Figure \ref{fig:results}) shows another interesting trend. Improvements in the CHANGE group happen faster than the DC group: most of the improvement for CHANGE occurs by the one-month survey, while improvements in the DC group are not fully realized until month three. Fast results are important for two reasons. First, rapid adoption of protective behaviors helps to immediately curtail transmission in a high-risk population. Second, high transience among YEH means that a non-negligible portion of youth will have left the center by the time a three-month intervention is completed. We conclude that the AI-augmented intervention implemented with CHANGE has substantial advantages over an intervention where peer leaders are selected with the standard DC method.

\section{Discussion}

This project provides evidence that AI methods can be used to improve the effectiveness of social network interventions in public health: significant reductions in HIV risk behaviors were observed in groups where our CHANGE method was used to plan the intervention, with no significant changes in behavior when the status quo method (selecting high degree nodes) was employed. More broadly, we hope that our experiences over the course of the project can provide generalizable lessons about how AI research can be successfully employed for social good. There have been recent attempts by others to synthesize principles for AI for Social Good research \cite{floridi2020design,tomavsev2020ai}. We offer a complementary perspective shaped by the process of deploying a specific community-level intervention. In particular, existing discussions of best practice often focus in large part on ethics, data privacy, and building trust with stakeholders. While such considerations are indispensable, it is also important for the research community to investigate the on-the-ground components of developing and deploying an impactful intervention. We highlight five points. 

First, the starting point was to listen to domain experts and understand where in the problem domain AI could be most impactful. We did not approach this project with a preexisting intention to apply influence maximization to the choice of peer leaders. Rather, this emerged organically from discussions between the AI and social work sides of the research team as a topic where an AI-augmented intervention was both technically feasible and likely to improve outcomes. \textit{Success is less likely when AI researchers start with a favored technique and search for an application. }

Second, data was overwhelmingly the bottleneck to the AI component of the intervention. Computational work on influence maximization to date had largely assumed a great deal of information would be known -- the structure of the graph, the model for information diffusion, etc. None of this information was in fact available for YEH (or would likely be available in other public health settings). Moreoever, gathering this data is itself time-consuming and costly, requiring unsustainable effort on the part of an agency wishing to deploy the intervention on their own. Much of the technical focus of the research consisted of finding ways to reduce the amount of data which needed to be gathered for the intervention to succeed. \textit{Finding ways to reduce or eliminate data needs through improved algorithm design is an important part of producing deployable AI interventions in a community health context.}

Third, simplicity is valuable. As an example, prior to developing CHANGE, we designed a much more theoretically sophisticated algorithm for collecting network data which enjoyed provable guarantees for certain families of graphs \cite{wilder2018maximizing}. However, it quickly became apparent that this algorithm would be difficult to deploy in practice because it required a large number of sequential queries (the node which is queried on step 1 determines the node who is to be queried on step 2, and so on). This was impractical in the context of a program working with YEH where any given youth may be difficult to find, interrupting the entire process. More generally, if the algorithm requires tight coupling with the outside world (many steps where information is input, the algorithm recommends a very specific action, more information is input, and so on) then there are more things that can go wrong which are not captured in the computational formalization of the problem. This poses a contrast to the way that simplicity is often operationalized in AI for social good work as either \textit{explainability} \cite{floridi2020design} or as \textit{methodological} simplicity \cite{tomavsev2020ai} (e.g., using well-developed techniques instead of a new algorithm). Both explainability and methodological simplicity are of course valuable in many settings but in our experience neither was first-order requirement: the algorithm can solve a complex optimization problem internally so long as the way that it interacts with the outside world is simple and robust.\textit{ We believe that this \textit{operational} simplicity is an under-emphasized design criterion for AI for Social Good. }

Fourth, smaller pilot tests were a valuable part of the project prior to embarking on a larger clinical trial. We conducted several such tests, each of which consisted of a deployment at a single drop-in center, in order to test earlier versions of our system \cite{wilder2018end,yadav2016using,yadav2017influence}. This helped reveal key issues which needed to be addressed. For example, we quickly discovered that a plan to collect network information via Facebook was not viable with this population and that manual collection of network data entailed a great deal of effort. We also quickly observed that peer leaders often did not attend the training, requiring on-the-fly adjustments over the course of the program. Addressing such issues was necessary to the success of the overall project (and turned out to provide much of the technical challenge involved). \textit{It would have been very difficult to identify these challenges without piloting algorithms in the actual environment where they will be used. }It was also helpful for computer scientists on the research team to be regularly present onsite during the pilot deployments to learn more about the environment and help coordinate the initial attempts at using the algorithm.

Fifth, community engagement and trust was essential to the success of the project. Beyond the research team, a number of stakeholders needed to be involved in the process. For example, we needed buy-in from each of the drop-in centers to conduct the study at the center, enroll their clients, and use their facilities. We regularly convened a community advisory board with representatives from each of the drop-in centers along with members of the research team to provide information about the study progress, explain the methods being used, and share information which could be helpful to other center activities. Just as critical as the center leadership though, were the youth themselves. We asked youth to disclose sensitive information, including their HIV risk behaviors and social contacts. Especially for the YEH population, which is less inclined than most to engage with authority figures, building trust is essential. We found two factors to be especially important in establishing this trust. First, the social work portion of the research team had deep roots in the community, having regularly offered services at these drop-in centers for the past ten years. Second, transparency about why information was being collected was critical. We observed substantially increased willingness to disclose information related to social contacts when researchers explained how this information would be used in the study (i.e., that a computer program would be used to select some people as peer leaders based on their contacts) than when such an explanation was not proactively given. \textit{A critical part of the peer change agent model is empowering youth to make a difference in their community, and this philosophy extends to the way that AI should be used in a community setting.}

Our hope is that this project provides one example towards a broader research agenda aiming at AI techniques which can be successfully used to improve health and equity within our communities. A great deal of work remains. Just within the context of social network intervention, future work should explore other intervention designs (e.g., interventions which attempt to modify network structure by fostering supportive relationships), methods for further reducing data requirements (e.g., by using administrative data to infer social connections), and more deeply investigate the relationship between information diffusion and behavioral change.  However, the results from this clinical trial provide evidence that AI can substantially improve the quality of services offered to the most vulnerable among us. 

\bibliographystyle{aaai}
\bibliography{main}

\end{document}